\documentclass[12pt]{article}
\usepackage{amssymb}

\def\be{\begin{equation}}
\def\ee{\end{equation}}
\def\bea{\begin{eqnarray}}
\def\eea{\end{eqnarray}}
\def\ba{\begin{array}}
\def\ea{\end{array}}

\def\Tr{\mathrm{Tr}}

\def\half{{\textstyle{\frac{1}{2}}}}

\catcode`\@=11
\def\@citex[#1]#2{%
\if@filesw \immediate \write \@auxout {\string \citation {#2}}\fi
\@tempcntb\m@ne \let\@h@ld\relax \def\@citea{}%
\@cite{%
  \@for \@citeb:=#2\do {%
    \@ifundefined {b@\@citeb}%
      {\@h@ld\@citea\@tempcntb\m@ne{\bf ?}%
      \@warning {Citation `\@citeb ' on page \thepage \space undefined}}%
      {\@tempcnta\@tempcntb \advance\@tempcnta\@ne%
      \@tempcntb\number\csname b@\@citeb \endcsname \relax%
      \ifnum\@tempcnta=\@tempcntb 
        \ifx\@h@ld\relax%
          \edef \@h@ld{\@citea\csname b@\@citeb\endcsname}%
        \else%
          \edef\@h@ld{\ifmmode{-}\else--\fi\csname b@\@citeb\endcsname}%
        \fi%
      \else
        \@h@ld\@citea\csname b@\@citeb \endcsname%
        \let\@h@ld\relax%
      \fi}%
    \def\@citea{,\penalty\@highpenalty\,}%
  }\@h@ld
}{#1}}

\def\@citeb#1#2{{[#1]\if@tempswa , #2\fi}}
\def\@citeu#1#2{{$^{#1}$\if@tempswa , #2\fi }}
\def\@citep#1#2{{#1\if@tempswa , #2\fi}}

\title{Noncritical $M$-theory and the Gross-Neveu model in $2+1$ dimensions}
\author{{\bf Anastasios C.~Petkou}\footnote{E-mail: petkou@physics.uoc.gr}\\
\em Department of Physics, \\
\em University of Crete,\\
\em 71003 Heraklion, GREECE.\\
\\
{\bf George Siopsis}\footnote{E-mail: siopsis@tennessee.edu}\\
\em Department of Physics
and Astronomy, \\
\em The University of Tennessee, Knoxville, \\
\em TN 37996 - 1200, USA.
}
\date{June 2006}
\begin{document}

\maketitle
\vspace{-5in}\hfill UTHET-05-0801\vspace{4.5in}

\abstract{We point out a non-trivial connection between the model proposed by Ho\v rava and Keeler as a candidate for noncritical $M$-theory
and the Gross-Neveu model with fermionic fields obeying periodic boundary conditions in $2+1$ dimensions.
Specifically, the vacuum energy of the former is identified with the large-$N$ free-energy of the latter up to an overall constant. This identification involves an appropriate analytic continuation of the subtraction point in noncritical $M$-theory, which is related
to the volume of the Liouville dimension.
We show how the world-sheet cosmological constant may be obtained from the
Gross-Neveu model. At its critical point, which is given in terms of the golden mean,
the values of the vacuum energy and of the cosmological constant are 4/5 and 2/5 of the corresponding values at infinite string coupling constant.

}
\newpage
\section{Introduction}

Non-relativistic fermi liquid models have been extensively used in the study of non-critical string theories in two dimensions \cite{bib1a,bib1b,bib1c,bib1d,bib1e,bib1f,bib1g,bib1h,bib1i,bib1j,bib1k,bib1l,bib1m,GinZinn}. In an interesting recent work \cite{bibhk} it has been suggested that fermi liquid model might also be used in the formulation of noncritical $M$-theory in three dimensions. The authors of \cite{bibhk} propose to identify the vacuum energy of two-dimensional non-relativistic fermions in an inverted harmonic oscillator potential with the energy of noncritical $M$-theory in $2+1$ dimensions. 

In this note, we show that the exact vacuum energy of the fermi liquid system calculated in \cite{bibhk} and the large-$N$ free energy of a three-dimensional Gross-Neveu model are given by remarkably similar expressions after an appropriate analytic
continuation of the subtraction point needed to define the density of states
in the fermi liquid system.
In order to make the identification, the fermionic fields of the model must be given
periodic boundary conditions along the compactified dimension.
This non-trivial relation points to an interesting correspondence between the two models.


\section{Noncritical $M$-theory}
\label{sec1}

We start with a discussion of the salient features of the $2+1$-dimensional model proposed by Ho\v rava and Keeler~\cite{bibhk}
aiming at casting the main results in a form appropriate for comparison with
the three-dimensional Gross-Neveu model.

In \cite{bibhk} noncritical M-Theory in $2+1$ dimensions can be described
by the double-scaling limit of non-relativistic fermionic fields $\psi (x^0, \bar x^1, \bar x^2)$ with action~\cite{bibhk}
\be I_M = \int dx^0 \int d^2\bar x \left( \psi^\dagger \partial_0 \psi
- \half \partial_i\psi^\dagger \partial^i \psi + V(\bar x) \psi^\dagger \psi \right)\,, \ee
where the potential is
\be V(\bar x) = \half \omega_0^2 \bar x^2 + \dots \,.\ee
The semiclassical density of states is obtained from the single particle
Hamiltonian
\be\label{eqspH} H(\bar p, \bar x) = \half \bar p^2 - V(\bar x)\,. \ee
The system is equivalent to a two-dimensional inverted harmonic oscillator
in the scaling limit of interest.
The density of states can be defined in terms of the resolvent
\be\label{eq1} \rho(\mu) \equiv \frac{1}{\pi} \Im \Tr \frac{1}{H+\mu-i\epsilon} = \frac{1}{\pi} \Re \int_0^\infty d\tau
\, e^{-i(\mu -i\epsilon)\tau} \Tr e^{-i\tau H} \ee
It may also be obtained by summing the type-0A contributions of all sectors with
integer values of the RR flux~\cite{bibhk}.

The partition function $Z\equiv \Tr e^{-i\tau H}$ appearing in~(\ref{eq1}) can be
derived from the partition function $Z_2$ of a normal
two-dimensional harmonic oscillator by continuing to imaginary frequency~\cite{bib1k}.
\footnote{It should be pointed out that one ought to exercise caution in performing such
an analytic continuation.
In the analogous two-dimensional case, justification for its validity may be found in~\cite{bibmoore}.
The calculation appies to the present three-dimensional case with little modification.}
Namely we consider
\be  Z_1 = \sum_{n=0}^\infty e^{-\tau (n+\half) \omega}\,, \ee
and we obtain
\be Z_2 = Z_1^2 = \frac{1}{4\sinh^2 \frac{\omega\tau}{2}} \,.\ee
To apply this result to the inverted two-dimensional harmonic oscillator, we rotate
$\omega \to -i\omega_0$ and $\tau \to i\tau$ and we deduce
\be Z \equiv \Tr e^{-i\tau H} = \frac{1}{4\sinh^2 \frac{\omega_0\tau}{2}}\,. \ee
Then the density of states~(\ref{eq1}) reads
\be \rho(\mu) = \frac{1}{4\pi} \Re \int_0^\infty d\tau
\, e^{-i\mu \tau} \, \frac{1}{\sinh^2 \frac{\omega_0\tau}{2}}\,. \ee
The integrand has a singularity as $\tau\to 0$ which can be regulated in a Pauli-Villars manner, by introducing a subtraction point $M$ and defining
\be \rho_M = \rho(\mu) - \rho(M)\,. \ee
The UV cutoff $M$ is related to the volume of the Liouville field in string theory.
To calculate the integral, we bring it into the form
\be \rho_M = \frac{1}{2\pi} \int_0^\infty d\tau
\, \frac{\sin^2 \frac{\mu\tau}{2} - \sin^2 \frac{M\tau}{2}}{\sinh^2 \frac{\omega_0\tau}{2}} \,.\ee
Making use of~\cite{bibgr}
\be \int_0^\infty dx \, \frac{\sin^2\beta x}{\sinh^2 \pi x} = \frac{\beta}{\pi (e^{2\beta} -1)} + \frac{\beta -1}{2\pi}\,, \ee
we obtain
\be\label{eq12} \rho_M = - \frac{\mu}{2\pi\omega_0^2}\, \coth \frac{\pi\mu}{\omega_0}
+ \frac{M}{2\pi\omega_0^2}\, \coth \frac{\pi M}{\omega_0}\,.
\ee
As $M\to\infty$, this is well approximated by
\be\label{eq13} \rho_M = - \frac{\mu}{2\pi\omega_0^2}\, \coth \frac{\pi\mu}{\omega_0}
+ \frac{M}{2\pi\omega_0^2}\, , \ee
in agreement with the result of~\cite{bibhk}.
Instead of~(\ref{eq13}), we shall use the more general form~(\ref{eq12}), which is valid for arbitrary values of $M$, in
order to compare with the Gross-Neveu model.

The vacuum energy is given by
\be F = \int^{-\mu} d\mu'\mu'\rho_M (\mu') \ee
Integrating, we obtain
\be
\label{free0}
F(\mu) = \frac{1}{2\pi^3\beta} \left\{ \frac{M\beta^3\mu^2}{8} \coth \frac{\beta M}{2} - \frac{\beta^3\mu^3}{12}
- \frac{\beta^2\mu^2}{2} \, \ln (1-e^{-\beta\mu})
+\beta\mu\,  \mathrm{Li}_2 (e^{-\beta\mu})
+ \mathrm{Li}_3 (e^{-\beta\mu}) \right\}
\ee
where we introduced the length scale
\be\label{eqMsc} \beta = \frac{2\pi}{\omega_0}\,, \ee
in order to facilitate comparison with the Gross-Neveu model.
The latter will be identified with the length of the finite dimension of the three-dimensional Euclidean space on which the Gross-Neveu model is defined.

To make contact with string theory one may introduce the conjugate variable
$\Delta$ (world-sheet cosmological constant) defined as
\be\label{eqLeg} dF = \mu d\Delta\,. \ee
Then the vacuum energy should be viewed as a function of $\Delta$ ($F = F[\mu(\Delta)]$).
One easily obtains
\be
\label{D0}
\Delta(\mu) = \frac{1}{4\pi^3} \left\{ \frac{M\beta^2\mu}{2} \coth \frac{\beta M}{2} - \frac{\beta^2\mu^2}{4}
- \beta\mu\, \ln (1-e^{-\beta\mu})
+ \mathrm{Li}_2 (e^{-\beta\mu}) \right\}\,.
\ee
In the large-$\mu$ limit, which may be thought of as corresponding to weak coupling, \footnote{Recall that in string theory  $\mu \sim g_s^{-1}$.} we have
\be F(\mu) \approx - \frac{\beta^2\mu^3}{24\pi^3} \ \ , \ \ \ \
\Delta(\mu) \approx - \frac{\beta^2\mu^2}{16\pi^3} \,,\ee
so that $F\sim \Delta^{3/2}$, to be compared with the string theory result
$F \sim \Delta^2 / \ln\Delta$.
Not only is the logarithm absent, we also have a string susceptibility exponent~\cite{bib1c,bib1d,bib1e,bib1f,bib1g} $\gamma_{\mathrm{str}} > 0$,
where $F\sim \Delta^{2-\gamma_{\mathrm{str}}}$,
indicating that we have crossed the string barrier of central charge $c\le 1$.

In the $\mu\to 0$ limit,  which analogously may be thought of as corresponding to strong coupling, we obtain
\be\label{eq20a} F(0) = \frac{1}{2\pi^3}  \frac{1}{\beta}\mathrm{Li}_3 (1) = \frac{\zeta(3)}{2\pi^3\beta}
\ \ , \ \ \ \ \Delta(0) = \frac{1}{4\pi^3} \mathrm{Li}_2 (1) = \frac{1}{24\pi}\,,
\ee
a behavior reminiscent of an asymptotically free theory~\cite{bib1k}.
For comparison, in string theory the vacuum energy approaches a finite constant
whereas $\Delta\to 0$.
Notice also that the UV regulator $M$ contributes in neither limit ($\mu\to 0$,
$\mu\to\infty$).

\section{The Gross-Neveu model}

Next, we consider the Gross-Neveu model in three Euclidean dimensions, one of which has finite length $\beta$.
The action is~\cite{Zinn-Justin,Rosenstein,Tassos2,Tassos3}
\be
\label{GNaction}
I_{GN} =-\int_0^\beta dx^0\int d^2\bar{x}\left(\bar{\psi}^a \slash\!\!\!\partial\psi^a +\frac{g}{2}(\bar{\psi}^a\psi^a)^2\right)\,,
\ee 
where $\bar{\psi}^a\,,\,\psi^a$ $(a=1,2,\dots,N)$  are here taken to be two-component Dirac fermions satisfying periodic boundary conditions along the finite dimension.\footnote{Periodic boundary conditions may be obtained by coupling the fermions to an
external $U(1)$ gauge potential which twists the antiperiodic boundary condition of the fermions to generically anyonic ones~\cite{Tassos3,Schreib}.} The $\gamma$-matrices are defined in terms of the Pauli matrices  as $\gamma^i=\sigma^i$ $(i=1,2)$ and $\gamma^0=\sigma^3$.

The path integral of the model gives the subtracted partition function as
\bea 
\label{ZGN1}
Z &=& e^{-\frac{N}{2}\beta [F(\beta)-F(\infty)]} \nonumber \\
&=& \int ({\cal D}\bar{\psi})({\cal D}\psi)({\cal D}\sigma) \exp\left\{ \int_0^\beta dx^0\int d^2\bar{x}\left(\bar{\psi}^a(\slash\!\!\!\partial+\sigma)\psi^a-\frac{1}{2g}\sigma^2\right)\right\} \nonumber \\
&=& \int ({\cal D}\sigma)\exp\left\{ N\left( \Tr_\beta\ln(\slash\!\!\!\partial +\sigma) -\frac{1}{2G}\int_0^\beta dx^0\int d^2\bar{x}\, \sigma^2\right)\right\}\,,
\eea
where $F$ denotes the free energy, $V_2$ denotes the  spatial volume and $G=gN$ is the rescaled coupling which is held fixed as $N\rightarrow\infty$ and $g\rightarrow 0$. In the second line of (\ref{ZGN1}) we have introduced the auxiliary scalar field $\sigma$ (with periodic boundary conditions) and in the third line we have integrated out the fermions.

We can evaluate the last path integral in (\ref{ZGN1}) by a saddle point expansion setting\footnote{We use on purpose the same notation $\mu$ here and in Section~\ref{sec1} to facilitate the comparison of results.} $\sigma=\mu +\lambda/\sqrt{N}$. Standard manipulations then give the leading-$N$ result
\bea
\label{ZGN2}
Z&=& \exp\left\{ N\beta V_2\left( \frac{1}{\beta}\sum_{n=-\infty}^{\infty}\int\frac{d^2\bar{p}}{(2\pi)^2}\ln(\bar{p}^2+\omega_n^2+\mu^2) -\frac{\mu^2}{2G}\right)\right\} \nonumber \\
&& \hspace{-1cm}\times\int({\cal D}\lambda) \exp\left\{ \sqrt{N} \int_0^\beta dx^0\int d^2\bar{x}\,\lambda(x)\left( \frac{2}{\beta} \sum_{n=-\infty}^{\infty}\int\frac{d^2\bar{p}}{(2\pi)^2}\frac{1}{\bar{p}^2 +\omega_n^2+\mu^2}-\frac{\mu}{G}\right)\right\},
\eea
where, due to the periodic  boundary conditions we have
\be\label{eqomn}\omega_n=\frac{2\pi n}{\beta} \ , \ \ \ \ n=0,\pm 1,\pm 2,\dots\ee  
From (\ref{ZGN2}) we read the large-$N$ free energy $F$ defined in~(\ref{ZGN1}),
\be
\label{freeGN}
\frac{F(\beta)}{V_2} = \frac{\mu^2}{2G}-\frac{1}{\beta}\sum_{n=-\infty}^{\infty}\int\frac{d^2\bar{p}}{(2\pi)^2}\ln(\bar{p}^2+\omega_n^2+\mu^2) +\int\frac{d^3p}{(2\pi)^3}\ln p^2\,,
\ee
and the gap equation ($\frac{\partial F}{\partial\mu} = 0$),
\bea
\label{gapGN}
\frac{\mu}{G}&=& \frac{2}{\beta}\sum_{n=-\infty}^{\infty}\int\frac{d^2\bar{p}}{(2\pi)^2}\frac{1}{\bar{p}^2 +\omega_n^2+\mu^2}\nonumber \\
&=& \frac{\mu}{G_*}-\frac{\mu^2}{2\pi} -\frac{\mu}{\pi\beta}\ln\left(1-e^{-\beta\mu}\right) +O(\Lambda^{-1})\,.
\eea
To arrive at the second line of (\ref{gapGN}) we used a large momentum cutoff $\Lambda$ and denoted
\be
\frac{1}{G_*}=\int^{\Lambda}\frac{d^3p}{(2\pi)^3}\frac{1}{p^2}\,.
\ee
The expression (\ref{freeGN}), up to an overall negative sign, coincides with the large-$N$ free energy density of the three-dimensional $O(N)$ vector model \cite{Tassos2} and the same applies also to the gap equation (\ref{gapGN}) up to the overall $\mu$ factor. In the $O(N)$ vector model, $-F(\beta)$ in (\ref{freeGN}) is interpreted as a physical free energy. In the case at hand, (\ref{freeGN}) can be a positive quantity and may be interpreted as energy. 

To proceed, let us define a critical parameter ${\cal M}$, with dimensions of mass as
\be
\frac{{\cal M}}{4\pi}=\frac{1}{G_*} - \frac{1}{G}\sim \Lambda^2(G-G_*)\,.
\ee
This will be kept finite (zero, negative or positive) in the scaling limit $\Lambda\rightarrow\infty$ and $G\rightarrow G_*$. ${\cal M}$ essentially quantifies the distance of $1/G$ from a critical value $1/G_*$ and determines the parameter $\mu$ by 
\be
\label{s0M}
2\sinh\frac{\mu\beta}{2}=e^{{\cal M}\beta /2}\,.
\ee
The parameter $\mu$ is the mass of the effective $\langle\lambda\lambda\rangle$ propagator, or alternatively it may be viewed as an inverse correlation length $\xi\sim 1/\mu$.
In the $O(N)$ vector model at zero temperature, it is an order parameter for the $O(N)\rightarrow O(N-1)$ symmetry breaking~\cite{Rosenstein}.

If we substitute $1/G$ from the gap equation~(\ref{gapGN}) into (\ref{freeGN}) and take the $\Lambda\rightarrow\infty$  limit, we find after some straightforward manipulations,
\be
\label{freeGN1}
F(\beta) = \frac{V_2}{\pi\beta^3} \left\{
- \frac{\beta^3\mu^3}{12}-\frac{\beta^2\mu^2}{2}\, \ln(1-e^{-\beta\mu}) +\beta\mu\, \mathrm{Li}_2(e^{-\beta\mu}) +\mathrm{Li}_3(e^{-\beta\mu})\right\}\,.
\ee
Crucial in obtaining (\ref{freeGN1}) was the fact that the linear divergence in $1/G_*$ canceled the linear divergence of the integral containing the logarithms in (\ref{freeGN}).

Comparing~(\ref{freeGN1}) with the result~(\ref{free0}) for the vacuum energy
in the Ho\v rava-Keeler model, we observe a remarkable non-trivial correspondence
provided we identify the length scale $\beta$ with the corresponding one in the
Ho\v rava-Keeler model (eq.~(\ref{eqMsc})) and the inverse correlation length $\mu$ with
the opposite of the fermi energy in the fermi liquid of the Ho\v rava-Keeler model.
The identification of length scales amounts to identifying the harmonic oscillator frequency $\omega_0$ in the Ho\v rava-Keeler model
with the fundamental frequency $\omega_1$ of the Gross-Neveu model (\ref{eqomn}) as
\be\label{eqo01}
\omega_0 =\omega_1 = \frac{2\pi}{\beta}\,.
\ee
The two expressions~(\ref{freeGN1}) of the Gross-Neveu model and (\ref{free0}) of the Ho\v rava-Keeler model differ in two respects.
\begin{itemize}
\item[{\em (a)}] In an overall constant dimensionless factor coming from the ratio $V_2/\beta^2$, and
\item[{\em (b)}] to obtain agreement, the subtraction point needed to define
the vacuum energy~(\ref{free0}) ought to be analytically continued to the
imaginary value
\be\label{eq20} M = \frac{(2n+1)\pi i}{\beta} \ \ , \ \ \ \ n\in\mathbb{Z} \ee
\end{itemize}
The free energy~(\ref{freeGN1}) should be viewed as a function of the critical parameter $\mathcal{M}$ (\ref{s0M}).
The latter is conjugate to the parameter $\mu^2$; at fixed temperature we have
\be \frac{dF}{V_2} = -\mu^2 \frac{d\mathcal{M}}{4\pi} \,,\ee
to be contrasted with the Legendre equation~(\ref{eqLeg}) defining the world-sheet
cosmological constant $\Delta$ in the Ho\v rava-Keeler model.
We may similarly obtain a two-dimensional ``cosmological constant'' in the
Gross-Neveu model starting from the expression~(\ref{freeGN1}) for the free energy.
Instead of introducing a Legendre transform for $\mu^2$, let us
define $\mathcal{D}$ by conjugating with respect to $\mu$, {\em i.e.,}
\be dF = \frac{\omega_0^2V_2}{2}\, \mu d\mathcal{D} \,.\ee
A short calculation yields
\be
\label{D}
\mathcal{D} = \frac{1}{4\pi^3} \left\{ \mathrm{Li}_2(e^{-\beta\mu}) -\frac{\beta^2\mu^2}{4}-\beta\mu\, \ln(1-e^{-\beta\mu}) \right\}\,,
\ee
which agrees with the the Ho\v rava-Keeler model parameter $\Delta$ (\ref{D0}) for the special
choice~(\ref{eq20}) of the subtraction point $M$.

In the limit $\mu\to 0$, we obtain
\be F(\beta) = \frac{V_2}{\pi\beta^3}\ \zeta(3) \ \ , \ \ \ \ \mathcal{D} = \frac{1}{24\pi} \,.\ee
Notice that $F(\beta)$ reduces to the free energy of a free fermion
and $\mathcal{D}$ to the strong-coupling expression for the cosmological constant~(\ref{eq20a}).

\section{Discussion}

We have shown that the fermi liquid model of noncritical $M$-theory introduced in~\cite{bibhk} yields an expression for the vacuum energy which agrees (up to an overall multiplicative constant) with the expression for
the free energy one obtains in the Gross-Neveu model, after analytical
continuation of the subtraction point $M$ (which is related to the
Liouville field volume) to one of the values~(\ref{eq20}).
To achieve agreement, we had to make two identifications:
\begin{itemize}
\item[{\em (i)}] the harmonic oscillator frequency $\omega_0$ was identified with the
fundamental frequency $\omega_1$ of the Gross-Neveu model (eq.~(\ref{eqo01})), and
\item[{\em (ii)}] the opposite of the fermi energy in the fermi liquid model was identified with the
inverse length scale $\mu$ (mass pole of the two-point function of the auxiliary field $\sigma$) of the Gross-Neveu model.
\end{itemize}
We could then obtain string dynamics in the Gross-Neveu model  as in the Ho\v rava-Keeler model by introducing a Legendre transform with respect to $\mu$.
However, one normally thinks of the critical parameter $\mathcal{M}$
in the Gross-Neveu model
in terms of a Legendre transform with respect to $\mu^2$, which yields the
gap eq.~(\ref{gapGN}).
For $\mathcal{M} > 0$ and fixed finite $\beta$, we cannot go to the $\mu\to 0$ limit.
Instead, we obtain a non-trivial critical point in the scaling limit $\Lambda\rightarrow\infty$ and $\mathcal{M}\to 0$, which corresponds to a non-trivial three-dimensional CFT \cite{Tassos2}.  
Solving the gap equation (\ref{gapGN}) for $\mathcal{M} = 0$, we obtain the critical
value
\be\label{eqgm} \mu \equiv \mu_* = \frac{2}{\beta}\, \ln\tau \ \ , \ \ \ \ \tau = \half (1+\sqrt 5)\,, \ee
given in terms of the golden mean.
The free energy at the critical point is
\be F(\beta;\mu_*) = \frac{4}{5}\, \frac{V_2}{\pi\beta^3}\ \zeta(3)\,. \ee
{\em i.e.,} 4/5 of its value at $\mu = 0$ (free fermion).
This is the result of non-trivial polylogarithm identities (see Appendix~\ref{appa}).
The same rational relationship holds for the vacumm energy in the Ho\v rava-Keeler model, if we allow $M$ to take on a value given by eq.~(\ref{eq20}).
We deduce from~(\ref{free0}),
\be\label{eqFst} F(\mu_*) = \frac{4}{5}\, F(0) \,.\ee
Moreover, the world-sheet cosmological constant at the critical point $\mu = \mu_*$ is related to its value at $\mu = 0$ (strong coupling limit) by
\be\label{eqDst} \Delta (\mu_*) = \frac{2}{5}\, \Delta(0)\,, \ee
which is obtained from~(\ref{D0}) for $M$ given by~(\ref{eq20}) and using dilogarithm identities (see Appendix~\ref{appa}).
It would be interesting to understand what two-dimensional critical model this
critical point might correspond to.

Finally, it would be of great interest to investigate the possibility of extending the observed correspondence between noncritical $M$-theory and the Gross-Neveu model in three dimensions to correlation functions. We hope to report on progress in this direction in the near future.

\section*{Acknowledgments}

The work of A.~C.~P.~is partially supported by the PYTHAGORAS II Research Program of the Greek Ministry of Higher Education.
G.~S.~is supported by the US Department of Energy under grant
DE-FG05-91ER40627.


\makeatletter
\@addtoreset{equation}{section}
\makeatother
\renewcommand{\theequation}{\thesection.\arabic{equation}}
                                                                                
\begin{appendix}
\section{Polylogarithm identities}\label{appa}

Here we present certain polylogarithm identities which
we used in order to derive eqs.~(\ref{eqFst}) and (\ref{eqDst}).
For more details see~\cite{Lewin}.

The $n$th-order polylogarithm is defined by
\be
{\rm Li}_n (x) = \sum_{k=1}^n {x^k\over k^n}
\ee
We have ${\rm Li}_n (1) = \zeta (n)$.

The dilogarithm obeys the following identities:
\be
\label{app1}
{\rm Li}_2 (x)+ {\rm Li}_2 (1-x) = {\rm Li}_2 (1)
-\ln x \ln (1-x)
\ee
\be
\label{app2}
{\rm Li}_2 (x)+ {\rm Li}_2 (-x) = {1\over 2}
{\rm Li}_2 (x^2)
\ee
\be
\label{app3}
{\rm Li}_2 (x) + {\rm Li}_2 \left( {-x\over 1-x}\right)
= - {1\over 2} \ln^2 (1-x)
\ee
The golden mean~(\ref{eqgm}) plays a special role in these identities.
Setting $x= 1/\tau$
in the first two identities, we obtain
\be
\label{app4}
{3\over 2} {\rm Li}_2 (1/\tau^2) - {\rm Li}_2 (-1/\tau)
= \mathrm{Li}_2 - {1\over 2} \ln\tau^2
\ee
Setting $x= 1/\tau^2$ in the third identity, we obtain
\be
\label{app5}
{\rm Li}_2 (1/\tau^2) - {\rm Li}_2 (-1/\tau) = - {1\over 8}
\ln^2 \tau^2
\ee
Combining Eqs.~(\ref{app4}) and (\ref{app5}), we finally
obtain
\be
\label{app8}
{\rm Li}_2 (1/\tau^2) = {2\over 5} {\rm Li}_2 (1) - {1\over 4}
\ln^2\tau^2
\ee
{\em i.e.}, the dilogarithm can be expressed in terms of
elementary functions at the special point $x=1/\tau^2$.
Eq.~(\ref{eqDst}) then easily follows from~(\ref{D0}) with $M$ given by eq.~(\ref{eq20}).

Similarly, for the trilogarithm, the following identities
hold:
\be
\label{app6}
{\rm Li}_3 (x)+ {\rm Li}_3 (1-x) + {\rm Li}_3 \left( {-x\over
    1-x}\right) = {\rm Li}_3 (1) 
+ {\rm Li}_2 (1) \ln (1-x)
-{1\over 2} \ln x \ln^2 (1-x) + {1\over 6} \ln^3 (1-x)
\ee
\be
\label{app7}
{\rm Li}_3 (x)+ {\rm Li}_3 (-x) = {1\over 4}
{\rm Li}_3 (x^2)
\ee
Again, the golden mean plays a special role. To see this, set
$x = 1/\tau^2$ in the first identity,
\be
{\rm Li}_3 (1/\tau^2)+ {\rm Li}_3 (1/\tau) + {\rm Li}_3 (-1/\tau) =
{\rm Li}_3 (1) 
+ {1\over 2} {\rm Li}_2 (1) \ln \tau^2
- {5\over 48} \ln^3 \tau^2
\ee
Using the second identity, we obtain
\be
{\rm Li}_3 (1/\tau^2) = {4\over 5} {\rm Li}_3 (1) - {2\over 5}
{\rm Li}_2 (1) \ln \tau^2 - {1\over 12} \ln^3 \tau^2
\ee
which, in view of Eq.~(\ref{app8}), can be written as
\be
\label{app9}
{\rm Li}_3 (1/\tau^2) + {\rm Li}_2 (1/\tau^2) \ln \tau^2 - {1\over 6}
\ln^3 \tau^2 
= {4\over 5} {\rm Li}_3 (1)
\ee
Eq.~(\ref{eqFst}) follows from eq.~(\ref{free0}) (with $M$ given by eq.~(\ref{eq20})) and eq.~(\ref{app9}).
\end{appendix}


\begin{thebibliography}{999}

\bibitem{bib1a} 
  M.~R.~Douglas, I.~R.~Klebanov, D.~Kutasov, J.~Maldacena, E.~Martinec and N.~Seiberg,
 ``A new hat for the c = 1 matrix $O(N)$ Vector Model,''
  arXiv:hep-th/0307195.
 \bibitem{bib1b} 
   T.~Takayanagi and N.~Toumbas,
  ``A matrix $O(N)$ Vector Model dual of type 0B string theory in two dimensions,''
  JHEP {\bf 0307}, 064 (2003)
  [arXiv:hep-th/0307083].
   
\bibitem{bib1c}
  P.~H.~Ginsparg and G.~W.~Moore,
  ``Lectures on 2-D gravity and 2-D string theory,''
  arXiv:hep-th/9304011.
 
\bibitem{bib1d} 
  I.~R.~Klebanov,
  ``String theory in two-dimensions,''
  arXiv:hep-th/9108019.
 \bibitem{bib1e} 
   J.~Polchinski,
  ``What is string theory?,''
  arXiv:hep-th/9411028.
 
\bibitem{bib1f} 
  S.~Alexandrov,
  ``Matrix quantum mechanics and two-dimensional string theory in  non-trivial
 backgrounds,''
  arXiv:hep-th/0311273.

\bibitem{bib1g} 
  Y.~Nakayama,
  ``Liouville field theory: A decade after the revolution,''
  Int.\ J.\ Mod.\ Phys.\ A {\bf 19}, 2771 (2004)
  [arXiv:hep-th/0402009].

\bibitem{bib1h} 
  E.~J.~Martinec,
  ``The annular report on non-critical string theory,''
  arXiv:hep-th/0305148.


\bibitem{bib1i}  
  J.~McGreevy and H.~L.~Verlinde,
  ``Strings from tachyons: The c = 1 matrix reloaded,''
  JHEP {\bf 0312}, 054 (2003)
  [arXiv:hep-th/0304224].

\bibitem{bib1j} 
  V.~A.~Kazakov and A.~A.~Migdal,
  ``Recent Progress In The Theory Of Noncritical Strings,''
  Nucl.\ Phys.\ B {\bf 311}, 171 (1988).
  
\bibitem{bib1k} 

  D.~J.~Gross and N.~Miljkovic,
  ``A Nonperturbative Solution Of D = 1 String Theory,''
  Phys.\ Lett.\ B {\bf 238}, 217 (1990).

\bibitem{bib1l} 
  D.~J.~Gross and I.~R.~Klebanov,
  ``One-Dimensional String Theory On A Circle,''
  Nucl.\ Phys.\ B {\bf 344}, 475 (1990).

\bibitem{bib1m} 

  E.~Brezin, V.~A.~Kazakov and A.~B.~Zamolodchikov,
  ``Scaling Violation In A Field Theory Of Closed Strings In One Physical
  Nucl.\ Phys.\ B {\bf 338}, 673 (1990).

\bibitem{GinZinn}
  P.~H.~Ginsparg and J.~Zinn-Justin,
  ``2-D Gravity + 1-D Matter,''
  Phys.\ Lett.\ B {\bf 240}, 333 (1990).
  
\bibitem{bibhk} 
  P.~Ho\v rava and C.~A.~Keeler,
  ``Noncritical M-theory in 2+1 dimensions as a nonrelativistic Fermi liquid,''
  arXiv:hep-th/0508024.

\bibitem{bibmoore} G.~Moore, ``Double-scaled field theory at $c=1$,'' Nucl.\ Phys.\ B {\bf 368}, 557 (1992).

\bibitem{bibgr} I.~S.~Gradshteyn and I.~M.~Ryzhik, {\sl Table of Integrals,
Series and Products,} Academic Press, 1994.

\bibitem{Zinn-Justin} J.~Zinn-Justin, {\it Quantum Field Theory and
Critical Phenomena}, 2nd ed.~Clarendon Press, Oxford, 1993.

\bibitem{Rosenstein} 

  B.~Rosenstein, B.~Warr and S.~H.~Park,
  ``Dynamical Symmetry Breaking In Four Fermi Interaction $O(N)$ Vector Models,''
  Phys.\ Rept.\  {\bf 205}, 59 (1991).

\bibitem{Tassos2} 

  A.~C.~Petkou and M.~B.~Silva Neto,
  ``On the free energy of three-dimensional CFTs and polylogarithms,''
  Phys.\ Lett.\ B {\bf 456}, 147 (1999)
  [arXiv:hep-th/9812166].


\bibitem{Tassos3}
  H.~R.~Christiansen, A.~C.~Petkou, M.~B.~Silva Neto and N.~D.~Vlachos,
 ``On the thermodynamics of the 2+1 dimensional Gross-Neveu model with complex chemical potential,''
  Phys.\ Rev.\ D {\bf 62} (2000) 025018
  [arXiv:hep-th/9911177].


\bibitem{Schreib}
  S.~Huang and B.~Schreiber,
 ``Statistical mechanics of relativistic anyon - like systems,''
  Nucl.\ Phys.\ B {\bf 426} (1994) 644.

\bibitem{bibnahm} 

  W.~Nahm, A.~Recknagel and M.~Terhoeven,
  ``Dilogarithm identities in conformal field theory,''
  Mod.\ Phys.\ Lett.\ A {\bf 8}, 1835 (1993)
  [arXiv:hep-th/9211034].

\bibitem{Lewin}
L. Lewin, {\it Polylogarithms and Associated Functions }, North
Holland, New York, 1981.



\end{thebibliography}
\end{document}